\documentclass[aps,prl,twocolumn,superscriptaddress]{revtex4}
\usepackage{graphicx}
\usepackage{amssymb}
\usepackage{amsmath}
\usepackage{graphicx}
\usepackage{epsfig}
\usepackage{color}
\usepackage{ulem}

\begin{document}

\title{Energy-scale cascade and correspondence between\\ the Mott and Kondo lattice physics}
\author{Danqing Hu}
\affiliation{Beijing National Laboratory for Condensed Matter Physics and Institute of Physics, Chinese Academy of Sciences, Beijing 100190, China}
\affiliation{University of Chinese Academy of Sciences, Beijing 100049, China}
\author{Ning-Hua Tong}
\affiliation{Department of Physics, Renmin University of China, Beijing 100872, China}
\author{Yi-feng Yang}
\email[]{yifeng@iphy.ac.cn}
\affiliation{Beijing National Laboratory for Condensed Matter Physics and Institute of Physics, Chinese Academy of Sciences, Beijing 100190, China}
\affiliation{University of Chinese Academy of Sciences, Beijing 100049, China}
\affiliation{Songshan Lake Materials Laboratory, Dongguan, Guangdong 523808, China}
\date{\today}

\begin{abstract}
We propose an energy-scale correspondence between the Mott physics and the Kondo lattice physics and construct a tentative phase diagram of their correlated electrons with two characteristic energy scales $\omega^*$ and $\Omega$ marking the upper boundary of a low-energy regime with well-developed long-range coherence and the lower boundary of localized spectral weight in the energy space, respectively. In between, there exists a crossover region with emergent but damped quasiparticle excitations. We argue that the presence of two separate energy scales is a generic property of correlated electrons on a lattice and reflects an intrinsic two-stage process of the quasiparticle dynamics to build up the lattice coherence. For the Hubbard model, they correspond to the kink and waterfall structures on the dispersion, while for the periodic Anderson model, they are associated with the indirect and direct hybridization gaps. Our work reveals a deep connection between the Mott and Kondo lattice physics and provides a basic ingredient for the study of many-body systems.
\end{abstract}

\maketitle

In correlated electron systems, a ``kink" denotes an abrupt slope change of the dispersive band and has often been regarded as a fingerprint of the coupling between electronic quasiparticles and collective bosonic excitations. It has been intensively studied in cuprates \cite{Lanzara2001Nature,He2001PRL,Shen2002PMB,Damascelli2003RMP,Meevasana2007,Armitage2010RMP}, where its appearance at an energy scale of around 30-90 meV has been attributed to spin fluctuations or phonons, which may potentially participate in the superconducting pairing. Different from these ``bosonic kinks" which represent the characteristic energy scales of the bosonic excitations, it has been shown later that kink may also arise purely due to electronic mechanism without involving bosonic excitations \cite{Byczuk2007NatPhys}. This ``electronic kink" was first seen in the Hubbard model (HM) within the framework of the dynamical mean-field theory (DMFT) \cite{Nekrasov2006} and then extended to other more complicated models \cite{Macridin2007PRL,Grete2011PRB,Kainz2012PRB,Greger2013PRL,Toschi2009,Toschi2010}. It might explain the so-called high energy kink observed over a wide energy range from 40 to 800 meV in cuprates and some other transition-metal oxides \cite{Yang2005PRL,Iwasawa2005,Yoshida2005,Ingle2005PRB,Valla2007,Graf2007,Xie2007PRL,Aizaki2012PRL}. 

Theoretically, the ``electronic kink" has been argued to mark an important energy scale of the Mott physics, namely the boundary of its Fermi liquid ground state \cite{Byczuk2007NatPhys}. Roughly speaking, the kink energy $\omega^*$ constrains a low-energy region of well-defined Landau quasiparticles and poses a threshold for the application of the Fermi liquid theory. A higher energy scale $\Omega$ has also been proposed as the boundary separating the many-body resonant state from the high-energy Hubbard bands. $\omega^*$ and $\Omega$ together define a cascade of key energy scales in the HM. Their discovery has since stimulated many debates concerning their true physical origin. Some suggested that the kink should still be of bosonic origin due to emergent collective spin fluctuations \cite{Rass2009}, while some associated it with an effective Kondo energy scale \cite{Held2013PRL}. A better understanding of its origin will deepen our knowledge of the many-body physics, but a consensus has yet to be reached. In particular, its fundamental importance and potential implications have not been fully recognized.

Here we point out their deep connection with the general physics of quasiparticle emergence and coherence and report a close correspondence between the energy scales of the Mott and Kondo lattice physics. While the two phenomena have mostly been studied separately in different families of correlated systems based on either the HM or the periodic Anderson model (PAM), there are increasing evidences supporting an intimate underlying connection \cite{Held2000PRL,Logan2016}. The Kondo lattice physics has been argued to be a special case of the (orbital-selective) Mott transition \cite{Medici2005,Pepin2007,Pepin2008,Vojta2010}, and the Mott physics in cuprates is itself a low-energy projection of the Kondo-Heisenberg model \cite{Zhang1988}. In the lately-discovered nickelate superconductors, both seem to exist and determine the properties of the ground state and low-energy excitations \cite{Zhang2020}. 

Our proposed correspondence is based on the simple observation that the two energy scales of the HM have the simple relationship $\omega^*\propto Z$ and $\Omega\propto {Z^{1/2}}$ with the renormalization factor Z \cite{Byczuk2007NatPhys,Bulla1999PB}. In comparison, the PAM also has two well-known energy scales, namely the indirect and direct hybridization gaps \cite{Coleman1984PRB,Auerbach1986PRL,Benlagra2011,Coleman2015}, which, in the mean-field approximation, are related to an effective hybridization $\tilde{V}$ through $\Delta_{\rm ind}\propto \tilde{V}^2$ and $\Delta_{\rm dir}\propto \tilde{V}$. This is reminiscent of $\omega^*$ and $\Omega$ in the HM. Since the hybridization gaps have well-defined physical meanings, it is natural to ask if this similarity can provide some insight on the physics of the HM or reveal certain generic features of both models. In this work, we will establish such a correspondence through systematic numerical studies and explore their origin and potential implications on the physics of correlated electrons. We will show that they represent two distinct groups of related energy scales and provide a good characterization of the energy boundaries separating the fully coherent and localized spectral weights with an itinerant but damped crossover region in between. This allows us to construct a tentative phase diagram in the energy space. We argue that the separation of the two energy scales reveals an instrinsic two-stage process of the quasiparticle dynamics for building up the long-range spatial coherence and represents a generic property of a typical class of correlated electron physics on a lattice represented by the HM and PAM. We note that the present work only concerns the paramagnetic Fermi liquid side of the HM phase diagram but for the whole spectrum beyond the Landau quasiparticle regime in the energy space. There is a difference between the energy domain and temperature domain. While the Fermi liquid is usually realized at low temperatures, the two energy scales should still be observed in the energy spectrum, reflecting the critical charge or spin dynamics of low-lying quasiparticle excitations.

\begin{figure}[t]
\includegraphics[width=0.49\textwidth]{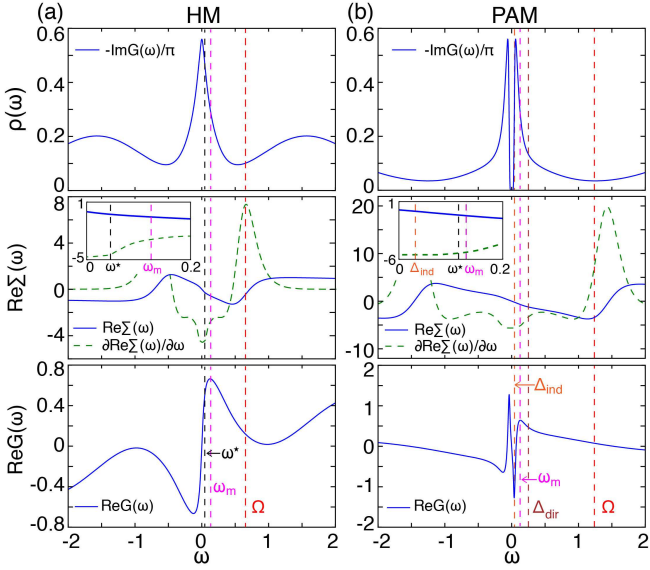}
\caption{Illustration of different energy scales identified from the $f$ electron DOS $\rho(\omega)=-{{\rm Im}}G(\omega)/\pi$, the real part of the $f$ electron self-energy ${\rm Re}\Sigma(\omega)$ and its slope ${\rm Re}\Sigma^\prime(\omega)$, and the real part of the Green function ${\rm Re}G(\omega)$ for (a) the HM with $U=3$ and (b) the PAM with $U=5.6$ and $V^2=0.4$. $\omega^*$ and $\omega_{\rm m}$ are estimated from the kink in ${\rm Re}\Sigma(\omega)$ and the maximum of ${\rm Re}G(\omega)$, respectively. For the HM, $\Omega$ is defined from the maximum of ${\rm Re}\Sigma^\prime(\omega)$. For the PAM, $\Delta_{\rm ind}$, $\Delta_{\rm dir}$, and $\Omega$ are obtained from the dispersion in Fig.~\ref{fig3}. The insets are the enlarged plots for the low energy scales: $\omega^*$, $\omega_{\rm m}$, and $\Delta_{\rm ind}$.}
\label{fig1}
\end{figure}

In both models, the Hamiltonians include two parts: $H=H_K+H_U$. The potential energy has the form, $H_U=U\sum_{i}\left( n_{i\uparrow}-\frac{1}{2}\right)\left( n_{i\downarrow}-\frac{1}{2}\right)$, where $U$ is the onsite Coulomb interaction and $n_{i\sigma}$ is the density operator of the local orbital. There are two general ways to delocalize the $f$ electrons. In the HM, one introduces a direct hopping between neighoring lattice sites, giving rise to a kinetic energy, $H_K=\sum_{k\sigma}\epsilon_{k}f_{k\sigma}^{\dag}f_{k\sigma}$, where $f^\dag_{k\sigma}(f_{k\sigma})$ are the creation (annihilation) operator of the $f$ electrons. In the PAM, the $f$ orbitals remain local and the delocalization is achieved by hybridizing with an additional conduction band such that $H_K=\sum_{k\sigma}\epsilon_{k}c_{k\sigma}^{\dag}c_{k\sigma}+V\sum_{i\sigma}(f_{i\sigma}^{\dag}c_{i\sigma}+{\rm H.c.})$, where $c^\dag_{k\sigma}$($c_{k\sigma}$) are the creation (annihilation) operator of the conduction electrons and $V$ is the bare hybridization parameter. For simplicity, we focus on the paramagnetic state in the half-filled one-band case with particle-hole symmetry and use the numerical renormalization group (NRG) \cite{Wilson1975, Bulla2008} as the impurity solver within the DMFT framework \cite{Metzner1989,Georges1992PRB,Georges1996RMP,Kotliar2006RMP}. The local self-energy of the $f$ electrons is calculated using $\Sigma(\omega)=UF(\omega)/G(\omega)$ with $G(\omega)=\langle\langle f_{i\sigma}; f_{i\sigma}^{\dag}\rangle\rangle_\omega$ and $F(\omega)=\langle\langle f_{i\sigma}(n_{i\overline{\sigma}}-1/2); f_{i\sigma}^{\dag}\rangle\rangle_\omega$ \cite{Bulla1998JPCM}. We then solve the DMFT self-consistent equation $G(\omega)=\int d\epsilon \rho_0(\epsilon)/[\omega-\epsilon-\Sigma(\omega)]$ for the HM and $G(\omega)=\int d\epsilon \rho_0(\epsilon)/[\omega-V^{2}/(\omega-\epsilon+i\eta)-\Sigma(\omega)]$ for the PAM \cite{Zhang1993PRL,Pruschke2000PRB}, where $\rho_0(\epsilon)=e^{-(\epsilon/t)^{2}}/(\sqrt{\pi}t)$ is the noninteracting density of states for a hypercubic lattice with the dimensionality $d \rightarrow \infty$. We set $t=1$ as the energy unit and choose the logarithmic discretization parameter $\Lambda=2$ and a total number of stored states $N_{\rm s}= 800$ for the NRG calculations. Our calculations are limited in the paramagnetic phase concerning the emergence and development of quasiparticle coherence. Magnetic instabilities are additional issues beyond the current work.

\begin{figure}[t]
\includegraphics[width=0.45\textwidth]{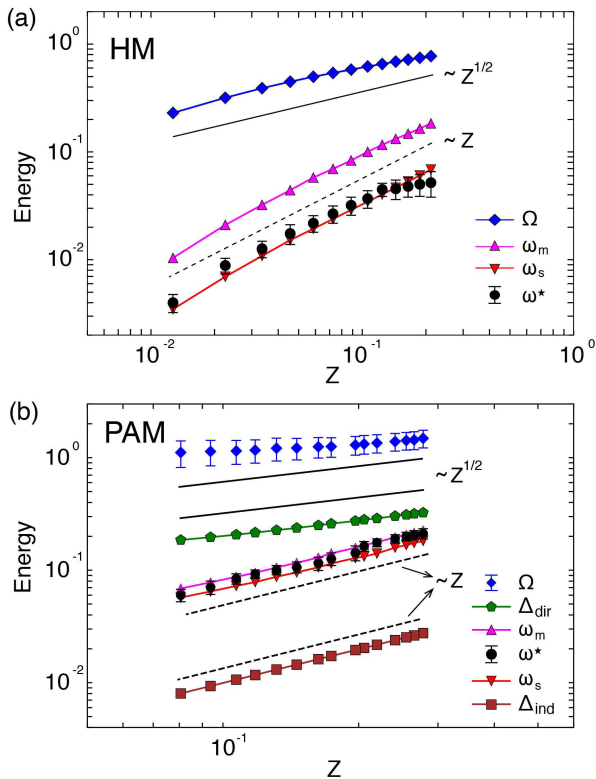}
\caption{The scaling relation of different energy scales with respect to the renormalization factor $Z$ for (a) the HM and (b) the PAM with tuning Coulomb interaction $U$. For the PAM, the hybridization strength is set to $V^2=0.4$.}
\label{fig2}
\end{figure}

Figure~\ref{fig1} illustrates how the different energy scales can be identified from the $f$ electron density of states (DOS) $\rho(\omega)=-{\rm Im} G(\omega)/\pi$, the real part of the self-energy  ${\rm Re}\Sigma(\omega)$ and its $\omega$-derivative ${\rm Re}\Sigma^\prime(\omega)$, and the real part of the Green function, ${\rm Re}G(\omega)$. For the HM, the DOS has a three-peak structure including two Hubbard peaks at higher energy and a quasiparticle peak around zero energy. The larger  scale $\Omega$ is defined here from the maximum of ${\rm Re}\Sigma^\prime(\omega)$. In Ref.~\cite{Byczuk2007NatPhys}, it was defined from the minimum of ${\rm Re}\Sigma(\omega)$, which, however, only follows the proposed scaling near the Mott transition with a small $Z$. Nonetheless, both reflect the same physics and separate roughly the quasiparticle peak and the Hubbard peaks in the DOS. The lower scale $\omega^*$ can be defined from an additional slope change or the kink in ${\rm Re}\Sigma(\omega)$ that constrains a low-energy region for quasiparticle excitations. A different scale $\omega_{\rm m}$ may be introduced from the maximum of ${\rm Re}G(\omega)$. $\omega^*$ and $\omega_{\rm m}$ are of the same origin and related through $\omega^*\approx(\sqrt{2}-1)\omega_{\rm m}$ for the HM \cite{Byczuk2007NatPhys}. A third definition may be found from the maximum ($\omega_s$) of the local spin susceptibility as discussed later in Fig.~\ref{fig4}. The relationship of these scales can be established if we tune the local Coulomb interaction $U$ and plot their variations with the renormalization factor, $Z^{-1}=1-{\rm Re}\Sigma^\prime(0)$. The results are summarized in Fig.~\ref{fig2}(a). We see that $\omega^*\propto\omega_m\propto Z$ and $\Omega\propto Z^{1/2}$ over a wide range of $Z$. Thus, $\omega^*$ and $\Omega$ represent two distinct groups of related energy scales. The different definitions originate from the crossover nature of the underlying physics. 

\begin{figure}[t]
\includegraphics[width=0.5\textwidth]{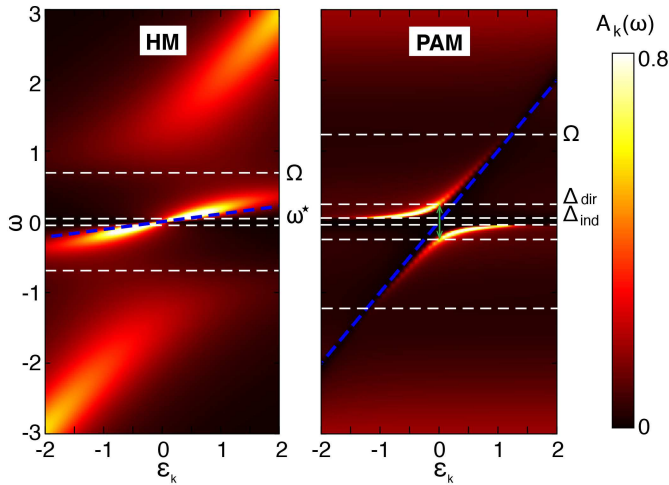}
\caption{Different energy scales on the intensity plot of the resolved spectral functions for the HM and PAM. The background colors reflect the magnitude of the $f$ electron spectral weight $A_{\rm k}(\omega)$. The blue dashed lines denote the dispersion $Z\epsilon_k$ for the HM and $\epsilon_{k}$ for the PAM. The parameters are the same as in Fig.~\ref{fig1}. For the PAM, the localized flat $f$ bands are outside the plotted energy window.}
\label{fig3}
\end{figure}

\begin{figure}[t]
\includegraphics[width=0.46\textwidth]{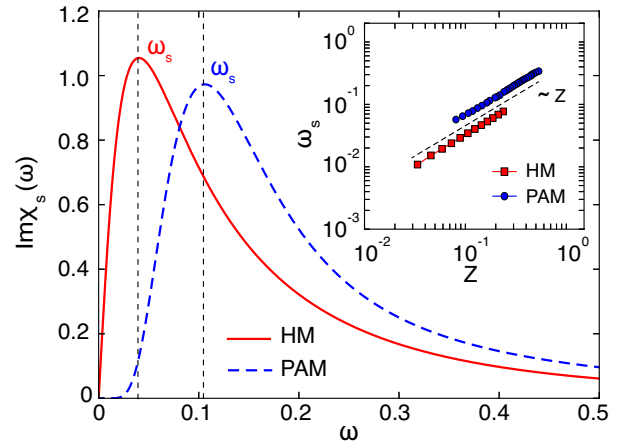}
\caption{The imaginary part of the local spin susceptibility, showing a maximum at $\omega_s$ for both models. The parameters are the same as in Fig.~\ref{fig1}. The inset plots $\omega_s$ as functions of $Z$ with $V^2=0.4$ (PAM) and varying $U$. }
\label{fig4}
\end{figure}

For comparison, Fig.~\ref{fig1}(b) shows the energy scales in the PAM. The indirect hybridization gap $\Delta_{\rm ind}$ is determined by the DOS gap and coincides with the additional minimum of ${\rm Re}G(\omega)$, while the direct hybridization gap $\Delta_{\rm dir}$ cannot be clearly discerned in these quantities and is best seen in Fig.~\ref{fig3} from the spectral function, $A_k(\omega)=-\pi^{-1}{\rm Im}[\omega-\Sigma(\omega)-\frac{V^{2}}{\omega-\epsilon_k+i\eta}]^{-1}$. The plot implies two separated hybridization bands for the PAM. The indirect gap $\Delta_{\rm ind}$ is the energy difference between the bottom of the upper band and the top of the lower band and marks the boundary for the low-energy gap. Although the PAM is in a Kondo insulating state, the self-energy below $\Delta_{\rm ind}$ behaves as that of a Fermi liquid \cite{Pruschke2000PRB}, reflecting the establishment of full coherence on the Kondo lattice \cite{Hu2019,Qi2020}. By contrast, the direct hybridization gap $\Delta_{\rm dir}$ measures the minimal energy difference between the two hybridization bands, below which the spectra are governed by the itinerant $f$ character. The energy scales $\omega^*$ and  $\omega_{\rm m}$ can still be defined similarly as in the HM and contain the flat part of the hybridization bands. We obtain $\omega^*\propto\omega_{\rm m}\propto\Delta_{\rm ind}$. On the other hand, the maximum of ${\rm Re}\Sigma^\prime(\omega)$ changes slightly with varying $U$ and is no longer a meaningful quantity separating the boundary of the itinerant and localized regions. Rather, we find it is better to define the larger energy $\Omega$ from the deviation of the hybridization bands from the original conduction bands. In Fig.~\ref{fig3}, it is seen that $\Omega$ sets roughly the outmost boundary of the itinerant $f$ electron spectral weight. Both $\Omega$ and $\Delta_{\rm dir}$ reflect the separation of the hybridized (itinerant) and unhybridized (localized) spectral regions of the $f$ electrons in the energy space.

The scaling results for the PAM are summarized in Fig.~\ref{fig2}(b). Similarly, we have $\omega^*\propto\omega_{\rm m}\propto\Delta_{\rm ind}\propto Z$ and $\Omega\propto\Delta_{\rm dir}\propto Z^{1/2}$ over a wide range of $Z$. The obtained scaling relations link $\omega^*$ and $\Omega$ to $\Delta_{\rm ind}$ and $\Delta_{\rm dir}$ and suggest that they reflect similar physics in the PAM. However, unlike $\omega^*$ and $\Omega$ which are in reality hard to measure, the indirect and direct hybridization gaps have unambiguous physical meaning and can be directly probed in experiment. Their scaling relations in the PAM can be derived from the pole properties of the Green function, $G_k(\omega)=[\omega-\Sigma(\omega)-\frac{V^{2}}{\omega-\epsilon_k+i\eta}]^{-1}$. If we make the lowest order approximation, ${\rm Re}\Sigma(\omega)\approx (1-Z^{-1})\omega$, the poles are determined by $Z^{-1}\omega-\frac{V^2}{\omega-\epsilon_k}=0$. The direct gap is defined as the energy differences of the two poles at $\epsilon_k=0$, while the indirect gap is the energy difference at the band edges ($\epsilon_k=\pm D$). We have $\Delta_{\rm dir}=Z^{1/2}V$ and $\Delta_{\rm ind}=ZV^2/D$, confirming their relationship with $Z$. The relative deviation of the hybridization bands from the conduction band may be estimated to be $\delta=(\omega-\epsilon_k)/\omega=ZV^2/\omega^2$. We have thus $\Omega\approx Z^{1/2}V/\delta^{1/2}$ for a chosen small cutoff $\delta$, at which the $f$ electron spectral weight on the dispersion is also reduced to the order of $Z\delta$. Since all these energy scales diminish at $V=0$, their presence reflects the delocalization of the $f$ electrons on the lattice driven by the hybridization with conduction electrons. In this respect, the larger scale $\Omega$ constrains the major spectral region of delocalized $f$ electrons, while $\omega^*$ puts a further restriction for well-established long-range coherence.

Similar analyses may help to understand $\omega^*$ and $\Omega$ in the HM. Figure~\ref{fig3} also plots the dispersion $A_k(\omega)=-\pi^{-1}{\rm Im}[\omega-\epsilon_k-\Sigma(\omega)]^{-1}$ of the HM. Indeed, we see that $\omega^*$ marks the boundary of well-defined Landau quasiparticles, above which the dispersion has a different renormalization factor, while $\Omega$ separates the low-energy many-body states and the high energy incoherent region, beyond which the $f$ electron weights are localized. The presence of both $\omega^*$ and $\Omega$ is a reflection of the special cascade structure of ${\rm Re}\Sigma^\prime(\omega)$, which, as shown in the middle panel of Fig.~\ref{fig1}(a), has a maximum at $\Omega$ but, with lowering energy, decreases first to a plateau (or local minimum) on the hillside before it eventually reaches the valley floor below $\omega^*$ and $\omega_{\rm m}$. Interestingly, $\Omega$ lies roughly at the waterfall structure connecting the quasiparticle and Hubbard bands \cite{Weber2008,Kohno2012}. This poses a question concerning the relation of the two properties. For this, we note that the waterfall is an abrupt change of the pole of the Green function with increasing $\epsilon_k$. It can only occur around the inflection point of $\omega-{\rm Re}\Sigma(\omega)$ if one wants to tentatively avoid multiple poles. This immediately implies ${\rm Re}\Sigma^{\prime\prime}(\Omega)\approx 0$ or a maximum (minimum) in ${\rm Re}\Sigma^{\prime}(\Omega)$. Thus, $\Omega$ defined here is indeed associated with the waterfall structure in the HM. We conclude that the two energy scales correspond to two features of the dispersion: $\omega^*$ for the kink and $\Omega$ for the waterfall, which also reflect the boundaries for the low-energy fully coherent region and the high-energy localized region. The fact that $\Omega$ connects two regions with opposite energy dependence in the self-energy implies $Z^{-1}\Omega \sim \Omega^{-1}$ or $\Omega\propto Z^{1/2}$, similar to that in the PAM. 

Further insight on $\omega^*$ may be obtained if we consider that $\Delta_{\rm ind}$ in the PAM is related to the spin screening, $\omega^*\propto\Delta_{\rm ind}\propto T_K$ where $T_K$ is the lattice Kondo temperature \cite{Jarrell1995PRB,Chen2016}. The effect of spin screening might be best seen from the local susceptibility, $\chi_s(\omega)=\langle\langle S_i^z;S_i^z\rangle\rangle_{\omega}$ with $S_i^z=\frac{1}{2}(n_{i\uparrow}-n_{i\downarrow})$ \cite{Shimizu2000JPSJ}. Figure~\ref{fig4} plots the imaginary part of the local spin susceptibilities for the HM and PAM using the same parameters as in Fig.~\ref{fig1}. Both exhibit a maximum at $\omega_s$, which, as shown in the inset, varies linearly with $Z$ as those of $\omega^*$ and $\Delta_{\rm ind}$. In fact, we find $\omega_s\approx \omega^*$ in both models (Fig.~\ref{fig3}). Away from half filling, this kink or hybridization energy scale $\omega^*$ still exists and retains the same relation with $\omega_s$ \cite{Grete2011PRB,Kainz2012PRB}. Below $\omega_s$ the local susceptibility is suppressed due to the spin screening. It is therefore natural to associate the kink with an effective spin screening scale as proposed in Ref.~\cite{Held2013PRL}. We should emphasize, however, that the presence of both $\omega^*$ and $\Omega$ is a property of the lattice but not in the usual single-impurity Kondo model where only the Kondo scale exists. This implies that the lattice feedback is crucial for separating $\omega^*$ and $\Omega$ and causing the cascade structure of ${\rm Re}\Sigma^\prime(\omega)$. Although DMFT does not contain explicit spatial correlations, it necessarily includes important lattice information through the self-consistent iterative procedure.

\begin{figure}[t]
\includegraphics[width=0.46\textwidth]{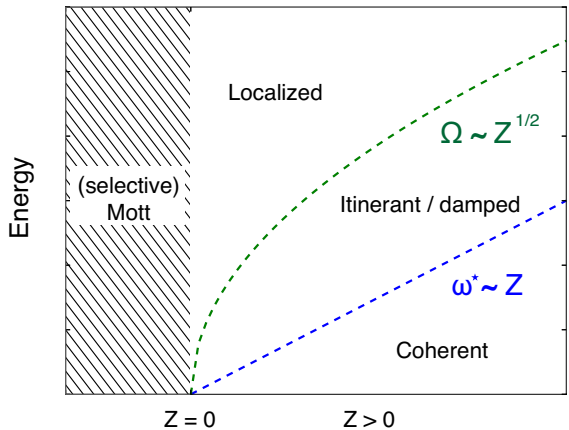}
\caption{A tentative phase diagram of the HM and PAM in the energy space as a function of the renormalization factor $Z$, showing different regions of correlated electron spectral weights. In the shadow region, the $f$ electrons become completely localized so that the HM is in a Mott state and the PAM is in an orbital-selective Mott state.}
\label{fig5}
\end{figure}

Thus, all energy scales can be classified into two categories, a lower one $\omega^*\propto Z$ and a higher one $\Omega\propto Z^{1/2}$, which separate the $f$ electron spectral weight into different regions of distinct properties. If we start from localized $f$ electrons, we may conclude that their spin screening and consequential delocalization is an energy-dependent process marked by the two scales. While $\Omega$ sets the outmost boundary for the many-body resonant states and covers the full energy range with delocalized $f$ electron spectral weight, $\omega^*$ is a lower boundary for quasiparticle excitations with well-established long-range coherence. This is common for both PAM and HM and may be best seen if we construct a tentative phase diagram in Fig.~\ref{fig5} in the energy space based on the properties of the $f$ electron spectra. For $Z>0$, the $\omega^*\propto Z$ line indicates a screening scale and marks the upper boundary of a well-developed coherent region irrespective of the Fermi liquid or Kondo insulating state, while $\Omega\propto {Z^{1/2}}$ marks the lower boundary of an incoherent region with localized $f$ spectral weight. In between, there exists a crossover region with itinerant but strongly damped excitations of the $f$ character. With decreasing $Z$ to zero, the system enters a Mott insulating state for the HM or an orbital-selective Mott state for the PAM, where the many-body resonance is suppressed and all $f$ electrons turn into fully localized magnetic moments. Since $Z$ is dimensionless, the scaling relations suggest constant prefactors roughly given by $D$ or $V^2/D$. Major correlation effects in both properties are encapsulated in the renormalization factor $Z$ (as a function of $U/D$) over a wide parameter region.

We remark again that the presence of the two energy scales represents two related but distinct features of the correlated electron physics. It is not just a matter of theory  but has real implications and observable consequences in experiment. To see this, we may consider the gradual delocalization of the $f$ electrons with lowering temperature. The two energy scales then turn into two temperature scales, as observed in recent pump-probe experiment on the heavy fermion compound CeCoIn$_5$ \cite{Qi2020}. Determinate Quantum Monte Carlo simulations (DQMC) taking into account spatial degree of freedom for the PAM have confirmed two distinct stages of hybridization dynamics, namely, the onset of precursor hybridization fluctuations and the eventual establishment of a long-range coherent (Kondo insulating) state \cite{Hu2019}. In the HM, the two energy scales are also expected to represent two succeeding stages of the (nonequilibrium) quasiparticle dynamics, namely, a Fermi liquid phase with well-established Landau quasiparticles and a higher crossover region with itinerant but damped quasiparticles (bad metal) \cite{Pruschke1995,Mravlje2011PRL,Deng2013PRL}. It might also be reflected in the temperature or time evolution of the doublon-holon excitations. Exact lattice simulations on the HM may help clarify this issue but are often limited by the numerical accuracy due to small lattice size. At even higher temperatures, the localized spectral weight might be thermally excited and contribute to physical properties, but whether or not (and how) the highly damped electrons can become Anderson localized requires more rigorous study \cite{Antipov2016}.

Along this line of thought, we anticipate that the existence and separation of two energy scales reflect a generic two-stage development of the electronic coherence and is an intrinsic property of correlated electrons. Before the long-range coherence is eventually established, there exists a precursor stage where the electrons become partially delocalized with damped quasiparticle excitations. On the other hand, the usual Mott transition also exists in other models such as the Falicov-Kimball model, the Ising-Kondo model, and the Hatsugai-Kohmoto model, which do not seem to possess the two energy scales. A closer inspection suggests that the Mott transitions in these models are different from that in the HM. In the Falicov-Kimball model and the Ising-Kondo model, the ground state near the transition is not a Fermi liquid due to ``disorder scattering"  \cite{Freericks2003,Yang2020}, while the Hatsugai-Kohmoto model lacks the so-called dynamical spectral weight transfer which is a key feature of the HM \cite{Yeo2019}. Thus the presence of two energy scales should not be viewed as a generic property of the Mott transition. Rather, they are associated with the emergence and establishment of coherent quasiparticles on the lattice, possibly induced by the spectral weight transfer from the high-energy localized part to the low-energy itinerant part. More elaborate studies are required to establish this important distinction.

To summarize, we report systematic analyses of the energy-scale cascade in the HM and the PAM and reveal a deep connection between the Mott and Kondo lattice physics. This allows us to construct a tentative phase diagram for correlated electrons based on two energy scales marking the upper boundary of the fully coherent low-energy states and the lower boundary of the incoherent regime with localized spectral weight in the energy space. For the HM, these correspond to the kink and waterfall structures on the dispersion. For the PAM, they are associated with the indirect and direct hybridization gaps. The separation of two energy scales is an intrinsic property of the lattice models and reflects a two-stage dynamical process to build up the lattice coherence of itinerant quasiparticles. Our work clarifies the origin of these energy scales and reveals a potentially basic and generic property for understanding the key of correlated electrons on a lattice.

This work was supported by the National Natural Science Foundation of China (NSFC Grant No. 11974397,  No. 11774401, No. 11974420), the National Key Research and Development Program of MOST of China (Grant No. 2017YFA0303103), and the Strategic Priority Research Program of the Chinese Academy of Sciences (Grant No. XDB33010100).

\end{document}